\documentclass[12pt]{article}
\pdfoutput=1

\usepackage{amsmath,amssymb,amscd}
\usepackage{listings}
\usepackage{caption}
\usepackage{dsfont}
\usepackage{slashed}
\usepackage{color}

\usepackage[pdftex]{graphicx}
\usepackage{epstopdf}
\usepackage{subfigure}
\usepackage{epsfig}
\usepackage{listings}
\usepackage{caption}
\usepackage{cite}

\usepackage{multirow}

\newcommand{\bea}{\begin{align}}
\newcommand{\eea}{\end{align}}
\newcommand{\beq}{\begin{equation}}
\newcommand{\eeq}{\end{equation}}
\newcommand{\nbea}{\begin{align*}}
\newcommand{\neea}{\end{align*}}
\newcommand{\nbeq}{\begin{equation*}}
\newcommand{\neeq}{\end{equation*}}
\newcommand{\bear}{\begin{eqnarray}}  
\newcommand{\eear}{\end{eqnarray}}  

\numberwithin{equation}{section}

\begin{document}

\begin{titlepage}

\pagestyle{empty}

\baselineskip=21pt
{\small
\rightline{Cavendish-HEP-17/02, DAMTP-2017-5}
}
\vskip 1in

\begin{center}

{\Large {\bf A Dynamical Weak Scale from Inflation}}

\vskip 0.5in

{\bf Tevong~You}

\vskip 0.3in

{\small {\it
{DAMTP, University of Cambridge, Wilberforce Road, Cambridge, CB3 0WA, UK; \\
Cavendish Laboratory, University of Cambridge, J.J. Thomson Avenue, \\ 
\vspace{-0.25cm}
Cambridge, CB3 0HE, UK}
}}

\vskip 0.75in

{\bf Abstract}

\end{center}

\baselineskip=18pt \noindent


{\small

Dynamical scanning of the Higgs mass by an axion-like particle during inflation may provide a cosmological component to explaining part of the hierarchy problem. We propose a novel interplay of this cosmological relaxation mechanism with inflation, whereby the backreaction of the Higgs vacuum expectation value near the weak scale causes inflation to end. As Hubble drops, the relaxion's dissipative friction increases relative to Hubble and slows it down enough to be trapped by the barriers of its periodic potential. Such a scenario raises the natural cut-off of the theory up to $\sim 10^{10}$ GeV, while maintaining a minimal relaxion sector without having to introduce additional scanning scalars or new physics coincidentally close to the weak scale.

}


\vskip 1in

{\small
\leftline{January 2017}
}

\end{titlepage}

\newpage


\section{Introduction}

Over the decades since its inception, the issue of naturalness of the Higgs mass has motivated a plethora of models in which this fine-tuning problem is considered solved in theory. However, an inevitable consequence of these theories is that fundamentally new physics should lie close to the TeV scale, and in light of null experimental searches at the LHC the Higgs naturalness issue is in danger of reverting back to being an unsolved open problem. Indeed, should null results persist in the search for a naturalising new sector at higher energies and precision, this may well prove to be our ``Michelson-Morley'' moment of the $21^\text{st}$ century. It is therefore worth reconsidering approaches to naturalness from as many different viewpoints as possible. 

Such an alternative approach has recently been put forward by GKR~\cite{GKR}~\footnote{This was inspired by Abbott's model originally proposed in the context of the cosmological constant~\cite{abbott}. We have nothing more to say about the cosmological constant here, though it may indeed turn out to be closely related to the Higgs mass fine-tuning problem~\cite{arvanitakietal}.}. Here the absence of new physics at the TeV scale is made compatible with a naturally light Higgs mass (relative to a large effective theory cut-off) as a result of cosmological evolution in the early universe.  In this work we will propose an alternative mechanism that relates the hierarchy ``weak scale/cut-off'', to a hierarchy in cosmological times ``$\Delta t_{Inf}/\Delta t_{RH}$'', where the former is the time change during inflation, and the latter the time change after reheating until the Hubble parameter falls below a certain threshold.  Throughout, we will make use of ingredients that are very familiar from previous work on weak-scale relaxation~\footnote{See Refs.~\cite{CHAIN, hardy, patilschwaller, antipinredi,  jaekeletal, guptaetal, naturalheavysusy, mirror, naturalrelaxation, koreanclockwork, kaplanrattazzi, dichiaraetal, ibanezetal, Nrelaxion, baryogenesisrelaxation, fowlieetal, doublescanningsusy, kobayashietal, hooktavares, choiim, flackeetal, mcallisteretal, relaxationafterreheating, higakietal} for some examples of recent studies of relaxation.}; however, as advertised, the underlying cosmological story is distinct.

As usual, let us imagine that there is a rolling scalar field, the relaxion, that scans the Higgs mass while it evolves down its scalar potential in the early Universe.  At some point the Higgs mass-squared will pass through zero.  In the usual relaxion models at this moment a technically natural backreaction of the Higgs vacuum expectation value $v$ will act upon the relaxion, causing it to halt, freezing $v$ at a small value.  Alternatively, in this work we propose that there is in fact no backreaction from the Higgs onto the relaxion.  Instead, the relaxion will continue rolling as before once the Higgs vacuum expectation value reaches a small value.  However, we will assume that the vacuum expectation value backreacts on the inflationary sector itself and will cause inflation to end.  We will provide two explicit examples for this dynamics.

Once inflation ends there is a short epoch while the inflaton rolls to the bottom of its potential and then oscillates and the Universe reheats.  While the Universe is cooling the Hubble value decreases, and finally the decreasing value of Hubble will backreact on the relaxion field by causing it to stop rolling, by some Hubble-dependent dynamics.

The key point is that the period of time in which the relaxion evolves during the inflationary epoch will typically be exponentially greater than the period of time between inflation ending and the Hubble parameter falling below some critical value.  Since the former epoch measures the change in the Higgs mass-squared between the cutoff and zero, $\Delta_{Inf} \sim M^2$, and the latter epoch measures the change in the Higgs mass squared from zero to a technically natural time depending on the Hubble-dependence of the relaxion potential, $\Delta_{RH} \ll M^2$, then a small Higgs vacuum expectation value becomes a prediction of the theory even though the Higgs itself does not play any role in halting the relaxion evolution.

In addition to the novel perspective on weak-scale relaxation, the advantage of decoupling the Higgs backreaction from the relaxion's trapping is that it simplifies considerably the relaxion sector of the theory. As we shall see, the minimal relaxion Lagrangian -- without any additional scalar fields or new electroweak-symmetry-breaking physics -- is sufficient for our model to naturally end up in a metastable vacuum with small weak scale.  

The paper is organised as follows. In Section~\ref{sec:relaxationsummary} we summarise the cosmological relaxation mechanism and how our model differs in its interplay with inflation. In Section~\ref{sec:inflation} we demonstrate examples of how inflation could end with a weak-scale vacuum expectation value for the Higgs. Constraints on the relaxion parameter space are listed in Section~\ref{sec:relaxionconstraints}. Finally, we conclude in Section~\ref{sec:conclusion}.

\section{Cosmological Relaxation}
\label{sec:relaxationsummary}

The relevant parts of the Lagrangian for the relaxion $\phi$, for an effective theory with a cut-off scale $M$, can be written as
\begin{equation}
\mathcal{L} \supset -\left(M^2 - g\phi\right)|h|^2 + gM^2\phi + \text{...} + \Lambda_G^4 \cos\left(\frac{\phi}{f_\phi}\right) - \frac{\alpha_D}{f_D}\phi F_{\mu\nu}\tilde{F}^{\mu\nu} \, ,
\label{eq:relaxionlagrangian}
\end{equation}
where $\Lambda_G$ is the strong coupling condensation scale of some gauge group $G$ and the ellipses denote terms higher order in $g\phi$ of the explicit shift symmetry-breaking potential $V(g\phi)$~\footnote{Including (or restricting to) the quadratic term does not modify the results while we are within the regime of validity of the effective field theory, $\phi \lesssim M^2/g$.}. The relaxion initially starts within a range $\phi \gtrsim M^2/g$ with a large positive effective Higgs mass squared that turns negative as $\phi$ rolls past the critical point $\phi_\text{crit.} \sim M^2/g$.  

In conventional relaxation models the barrier height of the periodic potential depends on the Higgs vacuum expectation value $v$ to some power $n$, 
\begin{equation}
\Lambda_G^4 \equiv \Lambda_G^{4-n} v^n \, .
\end{equation}
The dependence on $v$ provides a natural way to stop the scanning soon after the vacuum expectation value switches on and the barrier grows until it balances the linear slope when 
\begin{equation}
gM^2 \sim \frac{\Lambda_G^{4-n} v^n}{f_\phi} \, .
\end{equation}
The small value of $v$ relative to the cut-off is thus dynamically determined by this backreaction of the Higgs vacuum expectation value on the periodic potential that subsequently traps it. 

In the original GKR model~\cite{GKR}, $n=1$ and the gauge group $G$ was taken to be $SU(3)_c$ of QCD. Unfortunately this minimal realisation is excluded by the minimum of the relaxion generating too large an effective value of the QCD $\Theta$-angle. Alternatively $G$ could be a new strongly-coupled gauge group, but for $n=1$ this new physics has to be close to the weak scale. While more testable, this goes somewhat against the motivation of solving naturalness without such weak-scale physics and moreover introduces a coincidence problem~\cite{CHAIN}. Models with $n=2$ do not need to break electroweak symmetry in the new physics sector and can thus be decoupled to higher scales, but quantum corrections necessarily induce a barrier that prevents the relaxion from rolling. These initial barriers thus need to be relaxed by an additional scalar in a double-scanning mechanism~\cite{CHAIN, doublescanningsusy}. 

There are also $n=0$ models in which the periodic potential does not depend on the Higgs vacuum expectation value~\cite{hooktavares}. Note that the relaxion couples to a gauge field through the last term in Eq.~\ref{eq:relaxionlagrangian}, which may appear generically in theories involving axions as it respects a shift symmetry. In Ref.~\cite{hooktavares} this coupling was used as an alternative trapping mechanism through gauge boson dissipation~\footnote{See also Ref.~\cite{relaxationafterreheating} that uses gauge boson dissipation to prevent the relaxion from rolling too far after reheating.}. In this $n=0$ model the periodic potential barriers are always unsuppressed but the relaxion rolls with enough kinetic energy to overcome them. The Higgs mass squared starts out large and negative, so the vacuum expectation value is always on. It is then scanned down to low enough values that weak gauge bosons become light enough for dissipation to slow down the relaxion and trap it.

Such a mechanism has many advantages over requiring the periodic potential to depend on $v$. However, it also necessitates a UV completion preserving a specific coupling to the electroweak fields that allows, after electroweak symmetry breaking, a coupling to weak bosons but not to photons, 
\begin{equation*}
\mathcal{L} \supset -\frac{\phi}{f}\left( \alpha_{g^\prime} B_{\mu\nu}\tilde{B}^{\mu\nu}  - \alpha_g W^a_{\mu\nu}\tilde{W^a}^{\mu\nu} \right) \, ,
\end{equation*}
where $\alpha_{g^\prime}c_W^2 = \alpha_g s_W^2$, and $c_W \equiv \cos{\theta_W}$, $s_W \equiv \sin{\theta_W}$. This coupling pattern can be enforced by a LR symmetry, or appropriate group symmetry constructions~\cite{compositehiggs}, but a photon coupling could be reintroduced through higher-order corrections when the protecting symmetry is broken. Dissipation into photons would then trap the relaxion before reaching a small weak scale.

In the following we propose an alternative backreaction mechanism, where neither the periodic potential nor the dissipative gauge fields in the relaxion sector depend on the vacuum expectation value of the Higgs. Both are always there, with the relaxion having enough kinetic energy to roll over the barriers and the dissipation into ``dark'' gauge bosons being either sub-Hubble or providing the dominant friction for slow roll. In this scenario the Higgs mass squared is scanned down from large positive or negative values, depending on the realisation, and when the Higgs vacuum expectation value reaches the weak scale it triggers the end of inflation soon after. This causes Hubble to diminish, and as the relaxion's dissipation into gauge bosons increases relative to a falling Hubble it loses enough kinetic energy to become trapped. 

Such a mechanism provides a different physical viewpoint on the origin of the weak scale. Indeed, unlike the relaxion models considered so far, the backreaction is no longer directly responsible for trapping the relaxion. It is the Hubble scale that depends on $v$, with the transition from inflation to a radiation-dominated Hubble evolution determined by this vacuum expectation value, while the trapping occurs as Hubble drops below a critical value. The weak scale is thus linked to the ratio of the distance travelled by the relaxion during and after inflation, which is exponentially suppressed relative to the cut-off scale due to the exponentially long inflationary period.

\section{A weak-scale exit from inflation} 
\label{sec:inflation}

In this section we pursue in more detail the inflationary mechanism outlined above. For example, a simple and natural possibility for inflation to end at the weak scale is for the inflationary scale itself to be at the weak scale. Then the relaxion can play the role of the inflaton, while the Higgs acts as a waterfall field that ends inflation when electroweak symmetry is broken. In this case no additional inflaton sector is needed as our relaxion model already describes a hybrid inflation setup. This extremely minimal scenario is quite restrictive and would require a Hubble scale $H \sim 10^{-14}$ GeV. Such weak-scale hybrid inflation models have been considered in Refs.~\cite{weakscaleinflation}.  

We shall now discuss two other, more general, realisations of inflation that end when the Higgs acquires a weak-scale vacuum expectation value. The first relies on electroweak gauge field production to maintain slow-roll, while the second traps the inflaton through Higgs-dependent barriers whose height depends on the vacuum expectation value.

\subsection{Inflating with electroweak dissipation} 
\label{sec:inflatingwithelectroweakdissipation}

The production of gauge fields from a pseudoscalar field $\sigma$ originates from an axial gauge field coupling of the form
\begin{equation}
\mathcal{L} \supset -\frac{\alpha}{f}\sigma F_{\mu\nu}\tilde{F}^{\mu\nu} \, ,
\end{equation}
where $F_{\mu\nu}$ is the field strength for some abelian or non-abelian gauge field $A_\mu$ and $\tilde{F}_{\mu\nu} \equiv \epsilon^{\mu\nu\rho\sigma}F_{\rho\sigma}$ is its dual. This introduces an additional friction term in the equation of motion for $\sigma$ (see e.g. Refs.~\cite{anbersorbo, barnabyetal, lindeetal, dissipativeaxialinflation}), 
\begin{equation}
\sigma^{\prime\prime} + 2aH\sigma^\prime + a^2 V^\prime_\sigma(\sigma) = - \frac{a^2 \alpha}{f}\left< F_{\mu\nu}\tilde{F}^{\mu\nu} \right> \, ,
\end{equation}
where $a(\tau)$ is the scale factor of the homogeneous and isotropic metric, the primed derivatives of $\sigma$ are with respect to conformal time $\tau$, and $V^\prime_\sigma \equiv  \partial V / \partial\sigma$. 

Following Ref.~\cite{anbersorbo}, the equation of motion in an inflating universe for the two polarisations $A_\pm(k)$ can be written as 
\begin{equation}
\frac{d^2 A_\pm}{\tau^2} + \left(k^2 \pm \frac{k}{H \tau}\frac{ \alpha \dot{\phi}}{f} \right)A_\pm = 0 \, ,
\end{equation}
and the dot over $\phi$ denotes differentiation with respect to physical time $t$. This has the form of Whittaker's equation where the solution in terms of Whittaker functions yields the following equation of motion for $\sigma$,
\begin{equation}
\ddot{\sigma} + 3H\dot{\sigma} + V^\prime_\sigma(\sigma) = -I\frac{\alpha}{f}\left(\frac{H}{\xi}\right)^4 e^{2\pi\xi}  \, , 
\label{eq:eqmphi}
\end{equation}
with $I \sim 10^{-4}$, and we defined the dimensionless parameter
\begin{equation}
\xi \equiv \frac{\alpha}{2f}\frac{\dot{\sigma}}{H} \, .
\end{equation}
We shall be interested in the regime $\xi \gtrsim 1/(2\pi)$ for particle production to be significant. If the first two terms of the equation of motion Eq.~\ref{eq:eqmphi} are negligible compared to the potential slope and gauge field dissipation term then we obtain the following steady-state solution for $\xi$, 
\begin{equation}
\xi \simeq \frac{1}{2\pi}\log\left( \frac{9f M_p^4 |V^\prime_\sigma(\sigma)|}{I\alpha V^2(\sigma)} \right) \, .
\label{eq:xisol}
\end{equation}
Such an analytical solution has been shown to approximate well a more complete numerical analysis~\cite{dissipativeaxialinflation}. 

The assumption that the (modified) slow-roll parameter $\epsilon$ obeys the slow-roll condition $ \epsilon\ll 1,$ and that $V^\prime_\sigma > 3H\dot{\sigma}$ and $\ddot{\sigma}$, can be verified to hold when $\alpha \gtrsim \xi$~\cite{anbersorbo}. This also ensures $H^2 \simeq V(\sigma) / 3M_p^2$ such that the energy density of the gauge fields and kinetic energy $\dot{\sigma}^2$ does not dominate that of the inflaton potential. Since $\xi \gtrsim 1/(2\pi)$, and from Eq.~\ref{eq:xisol} we see that it only varies logarithmically under parameter changes, we shall always conservatively set $\alpha/\xi \sim \mathcal{O}(1)$. 

Dissipation allows for inflation on potentials that would otherwise be too steep for slow roll. Another advantage is that field excursions can be sub-Planckian since the number of e-folds of slow-rolling over a range $\Delta\sigma$ is given in this case by 
\begin{equation}
N_e \simeq \int_{\sigma_i}^{\sigma_f} H \frac{d\sigma}{\dot{\sigma}} \simeq \frac{\alpha}{2\xi} \frac{\Delta\sigma}{f} \, .
\label{eq:Nefolds}
\end{equation}
We shall see that depending on the parameters of our cosmological relaxation model the field excursions can indeed by sub-Planckian, although larger cut-offs will still require super-Planckian field excursions to obtain a large exponential numbers of e-foldings~\footnote{Ref.~\cite{anbersorbo} required $\alpha \sim \mathcal{O}(100)$ to obtain 45 e-foldings but this assumed a restricted field range $\Delta\sigma \lesssim \pi f$. Here we are employing an axion-like particle whose periodic potential coexists with an apparently non-compact field range. Such effective setups have been constructed in certain UV completions~\cite{koreanclockwork, kaplanrattazzi, clockworktheory}.}.

We now employ the gauge field production described above for $\sigma$ as the inflaton field. Our model of cosmological relaxation relies on ending inflation as the Higgs develops a vacuum expectation value, so if inflation is supported by electroweak gauge field dissipation this gives a natural way of ending slow-roll since the gauge boson masses are tied to the Higgs vacuum expectation value. 

 The inflaton $\sigma$ couples to to the electroweak gauge fields via the usual terms
\begin{equation}
\mathcal{L} \supset -\sigma\left( \frac{\alpha_B}{f_B} B_{\mu\nu}\tilde{B}^{\mu\nu}  + \frac{\alpha_W}{f_W} W^a_{\mu\nu}\tilde{W^a}^{\mu\nu} \right) \, , 
\end{equation}
which can be written in the mass eigenstate basis as 
\begin{equation}
\mathcal{L} \supset -\sigma\left(\frac{\alpha_W}{f_W}W^+_{\mu\nu}\tilde{W^-}^{\mu\nu} + \frac{\alpha_Z}{f_Z} Z_{\mu\nu}\tilde{Z}^{\mu\nu} + \frac{\alpha_\gamma}{f_\gamma} A_{\mu\nu}\tilde{A}^{\mu\nu} + \text{...} \right)  \, 
\end{equation}
where
\begin{equation}
\frac{\alpha_Z}{f_Z} \equiv \frac{\alpha_B s_W^2 f_W + \alpha_W c_W^2 f_B}{f_B f_W} \quad , \quad \frac{\alpha_\gamma}{f_\gamma} \equiv \frac{\alpha_B c_W^2 f_W + \alpha_W s_W^2 f_B}{f_B f_W} \, .
\end{equation}
Now while electroweak symmetry remains unbroken the inflaton dissipates into the massless $W^\pm$ and $Z$ bosons, as well as photons. The regime where such exponential particle production may dominate over Hubble is parametrised by
\begin{equation}
\xi_\sigma \equiv \frac{\alpha_W}{2f_W}\frac{\dot{\sigma}}{H} \gtrsim 1/(2\pi) \, ,
\end{equation}
and we assumed that the $W^\pm$ bosons provide the dominant dissipation channel, though the $Z$ may also contribute significantly. We do not require a photon coupling to be absent, only that it is unable to maintain slow roll inflation on its own.  

Maintaining slow-roll and self-consistency of the solution requires $\alpha_W \gtrsim \xi_\sigma$. Note that the solution for $\xi_\sigma$ in Eq.~\ref{eq:xisol} varies logarithmically and so is $\sim \mathcal{O}(1-10)$ for a general parameter space and potential $V(\sigma)$. In particular this means that the photon coupling can be fixed to have $\alpha_\gamma \lesssim \xi_\sigma$ without fine-tuning -- or at most a tuning of $\alpha_\gamma / \alpha_W \sim \mathcal{O}(10^{-1})$ -- ensuring that photo-dissipation does not maintain slow-roll when the weak gauge bosons gain mass and end inflation. 

In a thermal plasma at temperature $T$, non-Abelian gauge fields give rise to a thermal mass for the gauge bosons that can suppress particle production. However, such thermal effects will be diluted by inflation. If we treat the energy density of the gauge fields as a thermal bath we may estimate the temperature as 
\begin{equation}
T \simeq \left(\frac{\xi f V^\prime_\sigma}{C^* \alpha} \right)^{\frac{1}{4}} \gtrsim \left( \frac{\xi f H}{\sqrt{C^*}\alpha} \right)^\frac{1}{2} \, ,
\label{eq:gaugeT}
\end{equation}
where $C^* \equiv \frac{\pi^2}{30}g_\text{eff}$ is related to the effective number of degrees of freedom, and in the inequality we used the condition that dissipation dominates the equation of motion, $V^\prime_\sigma \gtrsim H\dot{\sigma} \sim \xi f H^2 / \alpha$. Note that Eq.~\ref{eq:gaugeT} assumes the production of gauge quanta thermalizes fast enough during inflation which may not trivially be the case. We shall therefore neglect the effects of thermal masses in our analysis and assume a standard cold inflation. 

The Hubble scale in this model must be no larger than around the weak scale. This is because the particle production threshold is approximately reached when 
\begin{equation}
\frac{\alpha_W}{f_W}\dot{\sigma} \sim m_V^* \quad \Rightarrow \quad H \sim \frac{m_V^*}{\xi_\sigma} \, , 
\end{equation}
where $m_V^* \lesssim v$ is the critical gauge boson mass. This therefore sets an upper limit on the Hubble scale during inflation to be $H \lesssim v$. When the particle production threshold is reached we assume the transition is sharp such that gauge boson production can no longer provide slow-roll as the weak scale is reached.

There is also a limit on how long inflation can last in these models, which is given by the number of e-foldings
\begin{equation}
N_e \simeq \frac{\Delta\sigma}{f_W} \, ,
\end{equation}
where $\Delta \sigma$ is the distance rolled by the inflaton. The field range of $\sigma$ is constrained by fixing Hubble, 
\begin{equation}
H^2 \simeq \frac{V(\sigma)}{3M_P^2} \, , 
\end{equation}
where a larger $\sigma$ must be compensated by a flatter slope $V^\prime(\sigma)$ to maintain a fixed value of $H$. On the other hand there is a lower bound on the slope,
\begin{equation}
V^\prime (\sigma) \gg 3H\dot{\sigma} \sim H^2 f_W \, ,
\end{equation}
in order for the dissipation dynamics to dominate over Hubble friction in the equations of motion. Putting all this together restricts the maximum number of e-foldings to 
\begin{equation}
N_e \simeq \frac{M_P^2}{f_W^2} \simeq 10^{18} \, \left(\frac{10^9 \, \text{GeV}}{f_W}\right)^2 \, ,
\end{equation}
where the decay constant must be $f_W \gtrsim 10^9$ GeV due to astrophysical constraints on electroweak interactions with axions in stars and supernovae~\footnote{For a review of constraints on axion couplings see e.g. Ref.~\cite{doddy} and references therein.}.

If the last decades of e-foldings correspond to those responsible for the currently observed CMB features then the axial gauge coupling may also give rise to many features in the cosmic microwave background, and the model parameters can then be constrained by generating the correct spectral index and density perturbations. For example, the scalar spectral index $n_s$ is given by~\cite{anbersorbo}
\begin{equation}
n_s = 1 + \frac{2}{\pi}\frac{f_W}{\alpha_W}\frac{V^{\prime\prime}(\sigma)}{V^\prime(\sigma)} \, .
\end{equation}
However the analytical solutions used here for the equations of motion may not be reliable towards the end of inflation where non-linear evolution can impact these observables~\cite{chengetal, cambridgepaper, dissipativeaxialinflation}. Moreover axial-gauge inflaton couplings typically induce large non-gaussianities that may further restrict the parameter space~\cite{barnabyetal, bugaevklimal}. A detailed numerical study is beyond the scope of this work where we are mainly interested in demonstrating the general mechanism. 

We remark that there is also the possibility of a brief second phase of inflation preceding a transition to our observed cosmology, as in hybrid inflation with a slow-rolling waterfall field. In the next section we shall give another example of a model with a second inflationary phase.

\subsection{Multi-natural inflation} 
\label{sec:QCDinflation}

\begin{figure}[h!]
\centerline{\includegraphics[height=6cm]{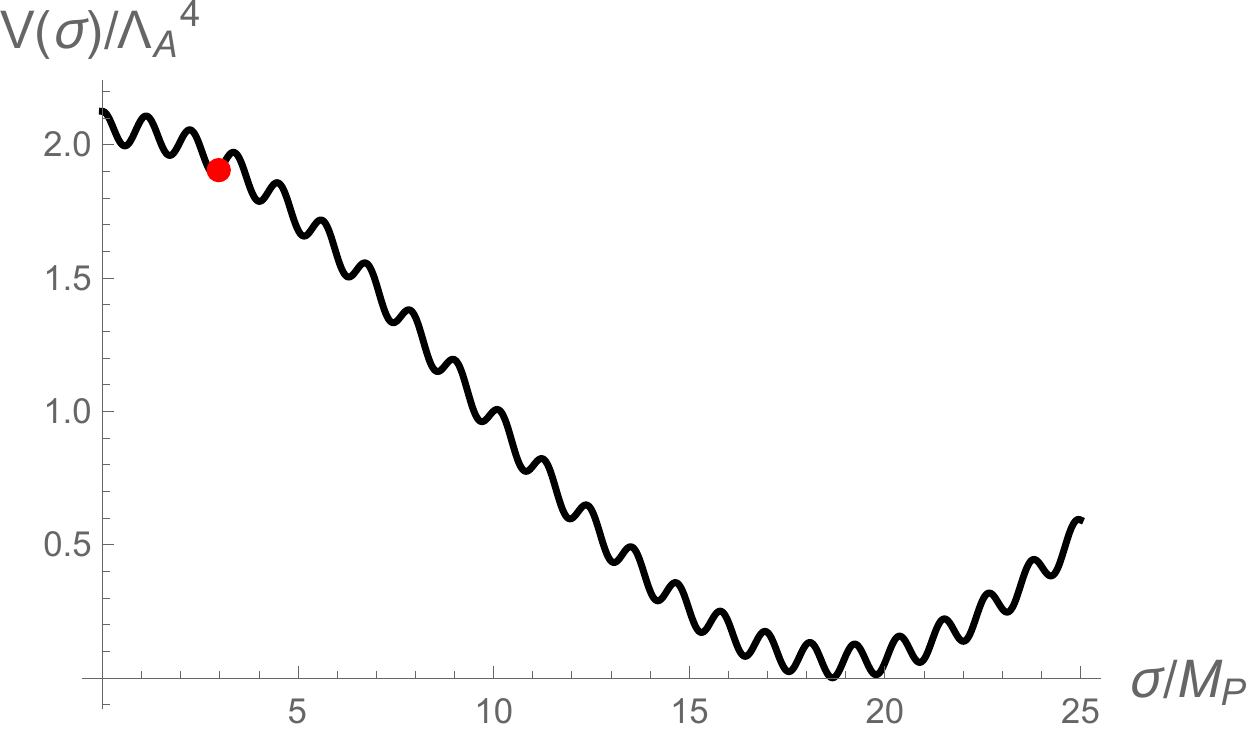}}
\caption{
{\it
Cartoon of the potential for our mechanism where the inflaton $\sigma$ denoted by a red dot is initially trapped. As the barriers of the smaller periodic potential depend on the Higgs vacuum expectation value, the inflaton is released towards the end of relaxation and begins a second phase of hilltop natural inflation. 
}} 
\label{fig:QCDpotentialplot} 
\end{figure}

In natural inflation~\cite{naturalinflation} an axion slow rolls on its periodic potential with large decay constant, and this may be modulated by an additional periodic potential in so-called multi-natural inflation models~\cite{multinaturalinflation}. If one of the periodic potentials depends on the Higgs vacuum expectation value then the inflaton can be trapped while the barrier height of that potential is initially very large. This maintains inflation while the relaxion scans the effective Higgs mass squared down to smaller values, until it reaches the point where the overall potential slope allows the inflaton to overcome the barriers. The mechanism is illustrated in Fig.~\ref{fig:QCDpotentialplot}. We outline here an example of how this may be achieved. 

The potential for multi-natural inflation can be written as
\begin{equation}
V(\sigma) = V_0 +\Lambda_A^4 \cos\left(\frac{\sigma}{f_A}\right) + \Lambda_B^4 \cos\left(\frac{\sigma}{f_B} + \delta\right) \, ,
\label{eq:Vsigma}
\end{equation}
where $f_A \gg f_B \gtrsim 10^9$ GeV (to avoid supernovae and stellar cooling constraints~\cite{doddy}), $V_0 \sim \Lambda_A^4$, and we set $\delta = 0$ for simplicity. We also have the hierarchy $\Lambda_A \gg \Lambda_B$, and the barrier height of the periodic potential $\Lambda_B^4$ is linearly proportional to the Higgs vacuum expectation value $v$. We note that as $v$ becomes large, as the strongly-coupled fermion mass responsible for the proportionality to $v$ decouples the condensate scale $\Lambda_B$ only grows logarithmically. This ensures that the potential is not dominated by $\Lambda_B$ when the Higgs vacuum expectation value starts out large with $v \sim M^2 / g$. As for $\Lambda_A$, its value fixes the energy scale of inflation and since Hubble must be less than the electroweak scale, $H \lesssim v$, we must set $\Lambda_A \lesssim 10^{10}$ GeV. 

As usual in small-field hilltop inflation, the inflaton is intially at the top of the hilltop close to the origin at $\sigma_i$. During relaxation it is prevented from rolling down the hilltop by a maximal barrier height $\Lambda_B$ while the Higgs vacuum expectation value is initially very large, $v \sim M^2 / g$. 

The lifetime for tunnelling out of this metastable local minimum must be exponentially long. We estimate this in the thin wall approximation as in Ref.~\cite{freeseetal} by the following probability for remaining in the false vacuum~\cite{guthweinberg}, 
\begin{equation}
P(t) \sim \text{exp}\left(-\frac{4\pi}{3}\beta Ht\right) \, ,
\end{equation}
where $\beta \equiv \Gamma / H^4$ is a dimensionless measure of the tunnelling rate per unit volume~\cite{tunnelling}, 
\begin{equation}
\Gamma \sim A e^{-S_E} \, ,
\end{equation}
the factor $A \sim \Lambda_A^4 f_B^2 / f_A^2$ is the difference in potential between the minima, and the Euclidean action $S_E$ is obtained by an integral over the barrier between the neighbouring minima of the potential,
\begin{equation}
S_E = \frac{27\pi^2 S_I^4}{2A^3} \quad , \quad S_I = \int_0^{2\pi f_B} \sqrt{2\left(V_0 + \Lambda_B^4 \cos{(\sigma/f_B)}  \right)} d\sigma \sim \Lambda_A^2 f_B \, .
\label{eq:SE}
\end{equation}
Since the tunnelling timescale is $t_\text{tunn.} \sim 3/4\pi\beta H$, the number of e-folds before tunnelling is roughly
\begin{equation}
N_e^\text{tunn.} \sim H t_\text{tunn.} \sim \frac{3}{4\pi}\frac{H^4}{\Gamma} \, .
\end{equation}
We see that it is mainly the size of $S_E$ that determines the exponent in the number of e-foldings before tunnelling. From Eq.~\ref{eq:SE} we find 
\begin{equation}
S_E \sim 10^{34} \left(\frac{10^{18} \text{ GeV}}{f_B}\right)^2\left(\frac{f_A}{10^{18} \text{ GeV}}\right)^6 \left(\frac{10^{10} \text{ GeV}}{\Lambda_A}\right)^4 \, . 
\end{equation}  
Though this is a rough estimate we may assume the tunnelling rate to be sufficiently suppressed during the exponentially long e-foldings of relaxation. 

As the relaxion scans $v$ down to small values at the weak scale, the relaxion is released at a critical value of $\Lambda_B \sim \Lambda_B^\text{crit.}$ given by the barrier slope condition:
 \begin{equation}
\Lambda_B^\text{crit.} \simeq \left(\frac{\sigma_i f_B\Lambda_A^4}{f_A^2}\right)^\frac{1}{4} \, .
 \end{equation}
At this point the $\Lambda_B$ periodic potential height is low enough to allow the inflaton $\sigma$ to overcome the barrier, allowing it to slow-roll towards larger field values in a second phase of hilltop natural inflation. 

We summarise here the basic formulae for cosmological parameters that characterise this phase of inflation. The slow-roll parameters are defined as
\begin{equation}
\epsilon(\sigma) \equiv \frac{1}{2} M_p^2\left( \frac{V^\prime(\sigma)}{V(\sigma)} \right)^2 \quad , \quad \eta(\sigma) \equiv M_p^2 \frac{V^{\prime\prime}(\sigma)}{V(\sigma)}  \, .
\label{eq:slowrollparams}
\end{equation}
The spectral index $n_s$ and tensor-to-scalar ratio $r$ are given by
\begin{equation}
n_s = 1 - 6\epsilon + 2\eta \quad , \quad r = 16\epsilon \, ,
\label{eq:ns}
\end{equation}
and are evaluated at the field value $\sigma^*$ when the relevant cosmological scale crosses the horizon. This is usually fixed by requiring a certain number of e-foldings before the end of inflation at $\sigma_e$,  
\begin{equation}
N^*_e = -\frac{1}{M_p^2}\int_{\sigma^*}^{\sigma_e} \frac{V(\sigma)}{V^\prime(\sigma)}d\sigma \, .
\label{eq:Nefolds}
\end{equation}

Applying these to the potential of Eq.~\ref{eq:Vsigma}, neglecting the small $\Lambda_B$ contribution, gives the following analytical expressions for the slow-roll parameters, 
\begin{align}
\epsilon(\sigma) &= \frac{1}{2}\left(\frac{M_P}{f_A}\frac{\sin(\sigma/f_A)}{\left(1 + \cos(\sigma/f_A)\right)}\right)^2 \, , \\
\eta(\sigma) &= -\left(\frac{M_P}{f_A}\right)^2\frac{\cos(\sigma/f_A)}{\left(1+\cos(\sigma/f_A)\right)} \, ,
\end{align}
and number of e-folds,
\begin{equation}
N_e = \frac{2f_A^2}{M_P^2}\log\left(\frac{\sin\left(\frac{\sigma_e}{2f_A}\right)}{\sin\left(\frac{\sigma}{2f_A}\right)} \right) \, .
\label{eq:Nefoldings}
\end{equation}
We see that $\eta \simeq - M_P^2 / 2f_A^2$ (when $\sigma \ll f_A$) requires a super-Planckian decay constant for the slow-roll condition $\eta \ll 1$ to hold from the start. 

Inflation ends when the slow-roll condition breaks down at $\epsilon, \eta \sim 1$. This happens at $\sigma_e < \pi f_A$ as the inflaton approaches the bottom of the potential, with $\sigma_e$ given by
\begin{equation}
\sigma_e \simeq 2f_A \arctan\left(\frac{\sqrt{2}f_A}{M_P}\right) \, .
\end{equation}
Using this we may solve Eq.~\ref{eq:Nefoldings} to obtain $\sigma^*$ for $N_e^*$ e-foldings before the end of inflation, 
\begin{equation}
\sigma^* \simeq 2f_A \arcsin\left(\frac{ e^{-\frac{N_e^* M_P^2}{2f_A^2}}}{\sqrt{1 + \frac{M_P^2}{2f_A^2}}} \right) \, .
\end{equation}
For $N_e^* \sim 60$ we then find that to obtain the observed spectral index $n_s(\sigma^*) \sim 0.96$~\cite{planck} requires $f_A \simeq  6 M_P$, and the corresponding tensor-to-scalar ratio is predicted to be $r \simeq 0.05$. The amplitude of the density perturbations can be characterised by the parameter
\begin{equation}
A_s  = \frac{1}{24\pi^2}\frac{1}{M_p^4}\frac{V}{\epsilon} \, ,
\label{eq:As}
\end{equation}
and is observed to be $A_s \sim 10^{-9}$~\cite{planck}. For $f_A \simeq 6 M_P$ this can only be obtained at a higher scale of inflation with $\Lambda_A \sim 10^{16}$ GeV. As mentioned previously we need a low scale of inflation for cosmological relaxation,  $\Lambda_A \lesssim 10^{10}$ GeV, so the density perturbations must be generated by fields other than the inflaton, as for example in curvaton models. 

The duration of the second phase of inflation from when it is released to the end of inflation depends on $\sigma_i$ and $\Lambda_B^\text{crit.}$. For example if $\Lambda_B$ is the QCD scale then the closer $\Lambda_B^\text{crit.}$ is to $\Lambda_\text{QCD}$ -- and hence the closer $\sigma_i$ is to the top of the hilltop -- the longer the second phase of slow-roll inflation lasts, up to a maximum of $\mathcal{O}(10^3)$ e-foldings for $f_A \sim 6M_P$. This therefore makes up a small fraction of the total relaxation period.

A multi-natural inflation model dependent on the Higgs vacuum expectation value serves to demonstrate how a second period of inflation after relaxation may help generate CMB observables, but other issues remain in this approach. If the QCD axion is used in this model then the potential minimum will in general generate a strong-CP problem~\footnote{See also Ref.~\cite{freeseetal} for another way of using the QCD axion as the inflaton. In their model a series of tunnelling rather than slow-rolling is responsible for inflation.}, whereas the QCD $\Theta$-angle is constrained by the neutron electric dipole moment to be $\sim 10^{-9}$~\cite{doddy}. A potential solution may be to ensure the phase shifts of the two potentials are aligned by discrete symmetries~\cite{hookCP} such that the $\Theta$-angle is still minimised by the multi-periodic potential.  

Another issue is that the small $\Lambda_B^\text{crit.}$ value requires the inflaton to start close to the origin where the slope is not too steep for it to remain trapped, while inflation would still occur had the inflaton been trapped elsewhere. To relax the fine-tuned initial condition requirement one could for example have a multi-natural inflation potential with a plateau on the hilltop such that the starting point is generic enough. There has also been some interesting comments regarding the possibility of tying the requirement of a high enough reheating temperature for generating matter-antimatter asymmetry to the relaxion initial conditions, providing a sharp and calculable anthropic explanation for such a setup~\cite{hooktavares}.   

We do not pursue a more realistic model as the general setup introduced here only illustrates how relaxation may be used to end an ``old inflation'' phase in a metastable vacuum, treating the multi-periodic potential as an effective setup. Such a potential may originate for example due to a coupling of a spontaneously-broken scalar to multiple fermions with different charges that condense at different strong coupling scales~\cite{multinaturalinflation}. In this case, the ratio of the different effective decay constants $f_A$ and $f_B$ are proportional to the ratio of the relative multiplicity of fermions and so should be close to each other, within a couple of orders of magnitude. However, other mechanisms that generate larger ratios may be possible. We leave the question of constructing a more realistic and UV-complete inflationary sector for future work.

\section{Constraints on the relaxion sector}
\label{sec:relaxionconstraints}

After inflation ends, Hubble falls below the inflationary Hubble scale $H_I$. At a critical value $H_c$ the relaxion is trapped when gauge field production becomes significant relative to Hubble and it dissipates kinetic energy efficiently. This threshold can be estimated by considering the equations of motion during a radiation-dominated universe,
\begin{equation}
\frac{d^2 A_\pm}{d\tau^2} + \left(k^2 + m_D^2 \mp \frac{a(\tau) k \alpha_D \dot{\phi}}{f_D}\right)A_\pm = 0 \, ,
\end{equation}
where the dark gauge boson may have a mass $m_D$. For a radiation-dominated universe the scale factor behaves as $a(t) \propto t^\frac{1}{2}$, or $a(\tau) \propto \frac{1}{2}\tau$ in conformal time, and Hubble goes as $H = \frac{1}{2} t^{-1}$. We may use the WKB approximation to the solution for the exponentially-growing positive helicity mode~\cite{relaxationafterreheating},
\begin{equation}
A_+ \simeq \frac{1}{\sqrt{2k}} e^{\int^\tau d\tau^\prime \sqrt{a(\tau^\prime)k a_D\dot{\phi}/f_D - k^2}} \, ,
\end{equation}
assuming a massless dark gauge boson for now. For a maximal gauge production mode $k \sim \frac{1}{2}a(\tau)\alpha_D\dot{\phi}/f_D$ we find that the exponent is $\gtrsim 1$ at a critical Hubble threshold 
\begin{equation}
H_c \simeq \sqrt{\frac{\alpha_D g M^2}{f_D}} \, .
\end{equation}
The choice of making the dark gauge boson massive or not allows more freedom to lower the critical Hubble scale, as when the gauge boson is massive the relaxion will then need to be rolling faster to trigger significant dissipation. Since the relaxion's gauge boson coupling is taken to be in a dark sector the parameter combination $\alpha_D / f_D$ and $m_D$ can always be fixed to the desired $H_c$. Next we turn to the general constraints on $H_c$, $g$ and $M$.

The relaxion must not roll further than the weak scale in the time that it takes to be trapped at $H_c$ from when inflation ends. This is the condition that $\int g\dot{\phi} dt < v^2$, assuming $\dot{\phi} \sim V^\prime(\phi) / H$ with $H \sim 1/t$, which sets a lower limit on $H_c$,
\begin{equation}
g^2M^2 \left(\frac{1}{H_c^2} - \frac{1}{H_I^2}\right) \lesssim v^2 \quad \Rightarrow \quad H_c \gtrsim \frac{gM}{v} \, .
\label{eq:rollpastvconstraint}
\end{equation}
On the right hand side of the arrow we have taken $H_I \gg H_c$ for the limit on $H_c$, though we will use this expression in general as a conservative constraint.

The slow-roll requirement $\epsilon(\phi) \sim M_P^2 / \phi^2 \ll 1$ with $\phi \sim M^2/g$ sets an upper limit on the slope whose steepness is parametrised by $g$, 
\begin{equation}
g \lesssim \frac{M^2}{M_P} \, .
\end{equation}
This turns out to be a weaker constraint than $g \lesssim H_c v / M$ from Eq.~\ref{eq:rollpastvconstraint}.

We may also obtain a bound on $M$ from requiring the inflaton energy density to dominate~\footnote{Note that this assumes $V_0$ is set according to the potential at $V(\phi \sim M^2 / g) \sim M^4$ which then determines the inflationary scale. This is the case for our two models of inflation in Section~\ref{sec:inflation}, but not in weak-scale hybrid inflation where the Higgs rolling to its electroweak-symmetry-breaking minimum triggers the end of inflation. In the latter, $V_0$ is tuned instead such that the inflationary scale is of the order the weak scale.},
\begin{equation}
H_I^2 \gtrsim \frac{V(\phi)}{3M_P^2} \quad \Rightarrow \quad M \lesssim \sqrt{H_I M_P} \simeq 10^{10} \, \text{GeV} \left(\frac{H_I}{10^2 \, \text{GeV}}\right)^\frac{1}{2} \, ,
\label{eq:inflatonenergydensitybound}
\end{equation}
so that using Eq.~\ref{eq:rollpastvconstraint} together with the cut-off scale $M$ that saturates the bound in Eq.~\ref{eq:inflatonenergydensitybound} fixes
\begin{equation}
g \lesssim \frac{H_c v}{\sqrt{H_I M_P}} \simeq 10^{-6} \, \text{GeV} \left(\frac{H_c}{10^2 \, \text{GeV}}\right)\left(\frac{10^2 \, \text{GeV}}{H_I}\right)^{\frac{1}{2}} \, .
\label{eq:g}
\end{equation}
Choosing $H_I$ and $H_c$ then fixes $M$ and $g$. 

A further constraint comes from classical rolling beating quantum fluctuations, $H_I < \dot{\phi}\Delta t$ within a Hubble time $\Delta t \sim 1 / H_I$, which gives
\begin{equation}
H_I^3 \lesssim gM^2 \, .
\end{equation}
Substituting $M$ and $g$ with Eqs.~\ref{eq:inflatonenergydensitybound} and \ref{eq:g} yields a lower bound on $H_c$, 
\begin{equation}
H_c \gtrsim \frac{1}{v}\left(\frac{H_I^5}{M_P}\right)^\frac{1}{2} \simeq 10^{-6} \, \text{GeV} \left(\frac{H_I}{10^2 \, \text{GeV}}\right)^\frac{5}{2} \, .
\label{eq:Hclimit}
\end{equation}

\begin{figure}[h!]
\centerline{\includegraphics[height=10cm]{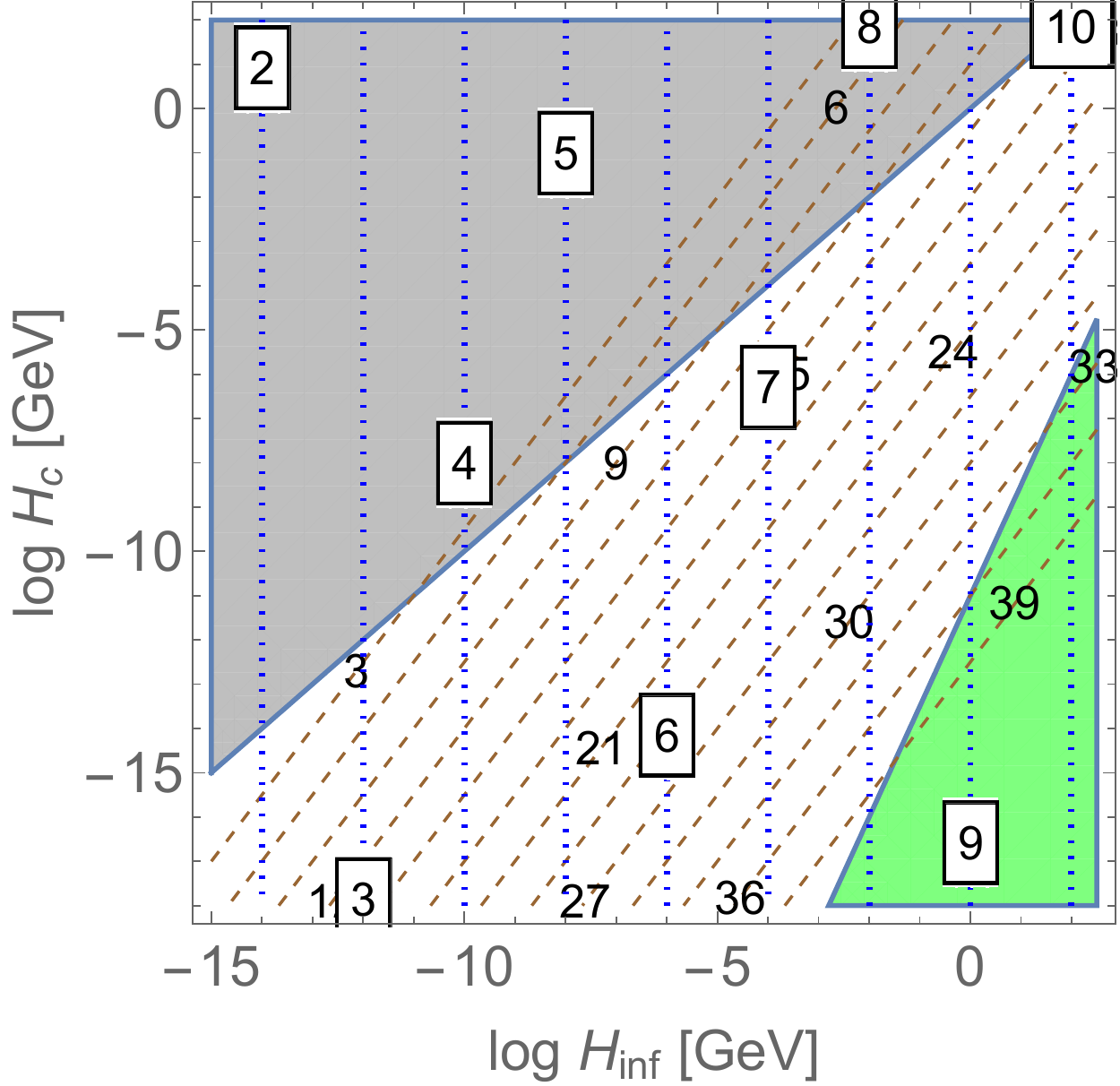}}
\caption{
{\it
Parameter space of the relaxion sector determined by the Hubble scale of inflation $H_\text{inf}$ vs the critical Hubble threshold $H_c$ at which the relaxion is trapped. The upper grey shaded region is excluded because the latter is restricted to $H_c \lesssim H_\text{inf}$, and the lower green shaded region is when $H_c$ is too small so the relaxion is trapped after rolling past the weak scale. The vertical blue dotted lines labelled by white rectangles denote the log of the maximum cut-off $M$ in GeV. The diagonal brown dashed lines are the log of the number of e-foldings, for the value of $g$ and $M$ that saturate the bounds in Eqs.~\ref{eq:rollpastvconstraint} and ~\ref{eq:inflatonenergydensitybound}.
}} 
\label{fig:HinfvsHc} 
\end{figure}

We now determine the relaxion trapping conditions. Before it is trapped, the size of the periodic potential barriers must be such that during slow-roll the relaxion's kinetic energy is large enough to go over the barriers, $\dot{\phi}^2 \gtrsim \Lambda_G^4$, so that 
\begin{equation}
\Lambda_G^2 \lesssim \frac{gM^2}{H_I} \, .
\end{equation}
In terms of the $g$ and $M$ saturating the bounds in Eqs.~\ref{eq:inflatonenergydensitybound} and \ref{eq:g} this is
\begin{equation}
 \Lambda_G \lesssim \left(\frac{M_P}{H_I}\right)^\frac{1}{4}\sqrt{H_c v}  \simeq 10^{6} \, \text{GeV} \left(\frac{10^2 \, \text{GeV}}{H_I}\right)^\frac{1}{4}\left(\frac{H_c}{10^2 \, \text{GeV}}\right)^\frac{1}{2} \, .
 \end{equation}
 After dissipating its kinetic energy the relaxion will remain trapped so long as the slope satisfies the multiple minima condition, 
\begin{equation}
 gM^2 \lesssim \Lambda_G^4 / f_\phi \, ,
\end{equation}
 which ensures the linear slope of the explicit breaking potential is not too steep relative to the periodic potential. This can be written, using $g$ and $M$ as previously, as an upper limit on $f_\phi$: 
 \begin{align}
f_\phi &\lesssim \frac{\Lambda_G^4}{H_c v \sqrt{H_I M_P}} \\
&\simeq 10^{10} \, \text{GeV} \left(\frac{\Lambda_G}{10^6 \text{ GeV}}\right)^4\left(\frac{10^2 \text{ GeV}}{H_c}\right)\left(\frac{10^2 \text{ GeV}}{H_I}\right)^\frac{1}{2} \, . \nonumber
 \end{align}
Finally, we must also ensure that scanning occurs through small enough steps of the periodic potential such that the difference between adjacent minima is not larger than the weak scale, $gf_\phi \lesssim v^2$, which restricts
\begin{equation}
f_\phi \lesssim \frac{v \sqrt{H_I M_P}}{H_c} \simeq 10^{10} \, \text{GeV} \left(\frac{H_I}{10^2 \, \text{GeV}}\right)^\frac{1}{2}\left(\frac{10^2 \, \text{GeV}}{H_c} \right) \, .
\end{equation}

The number of e-folds required during the cosmological relaxation phase of inflation is given by 
\begin{equation}
N_e \simeq \frac{H_I^2}{g^2} \simeq \frac{H_I^3 M_P}{H_c^2 v^2} \simeq 10^{16} \left(\frac{H_I}{10^2 \, \text{GeV}}\right)^3 \left(\frac{10^2 \, \text{GeV}}{H_c}\right)^2 \, .
\end{equation}
For example with a low-enough Hubble scale $H_I \sim H_c \sim 10^{-12}$ GeV and cut-off $M \sim \mathcal{O}(\text{TeV})$ we can have $N_e \sim \mathcal{O}(100)$ and a sub-Planckian field excursion. 

\begin{table}[h!]
\begin{center}
\begin{tabular}
{| c | c | c | c | c | c | c | c | c |}
\hline
& $M$ & $g$ & $H_I$ & $H_c$ & $N_e$ & $\Lambda_G$ & $f_\phi$ & $f_D/\alpha_D$ \\
\hline
$\sim$ [GeV] & $10^8$ & $10^{-11}$ & $10^{-2}$ & $10^{-5}$ & $10^{18}$ & $10^{3.5}$ & $10^{9}$ & $10^{15}$ \\
\hline
\end{tabular}
\end{center}
\caption{\it An example of typical parameter values that satisfy all the constraints listed for our relaxion model. $M$ is the effective theory cut-off, $g$ parametrises the explicit shift-symmetry-breaking slope, $H_I$ is the Hubble scale of inflation, $H_c$ the critical Hubble threshold below which the relaxion is trapped, $N_e$ the required e-foldings of inflation during relaxation, $\Lambda_G$ the trapping barrier height of the relaxion's periodic potential with period $2\pi f_\phi$, and $f_D$ the decay constant of the relaxion's axial gauge field coupling responsible for dissipation into dark gauge bosons.   }
\label{tab:benchmark}
\end{table}

A benchmark point that satisfies all the constraints outlined above is summarised in Table~\ref{tab:benchmark}~\footnote{Further cosmological and direct search constraints may be placed on general relaxion models, see for example Ref.~\cite{flackeetal}.}. The general parameter space can be characterised by the Hubble scale of inflation and the critical Hubble threshold, as shown in Fig.~\ref{fig:HinfvsHc}. The upper grey and lower green excluded regions are due to requiring $H_c \lesssim H_I$ and $H_c$ to satisfy Eq.~\ref{eq:rollpastvconstraint} respectively. This ensures the relaxion is trapped before it rolls past the weak scale after inflation ends. The vertical dotted blue lines are the upper limit on the cut-off $M$ in GeV and the diagonal dashed brown lines are the number of e-foldings during inflation, both on a logarithmic scale. As mentioned above we used the values for $g$ and $M$ that saturates their respective bounds. We see that the typical number of e-folds can vary anywhere between $\sim 10^3 - 10^{39}$. 

Note that we have considered the relaxion to be classically slow-rolling using Hubble friction, with negligible dissipation until Hubble drops below the critical threshold. Alternatively we could also have dark dissipation dominating at all times. In this case it may also slow-roll on a steep potential similarly to the case of the inflation model with gauge dissipation of Section \ref{sec:inflatingwithelectroweakdissipation}. It is straightforward to rederive the above relations using instead the field evolution expressions from Section \ref{sec:inflatingwithelectroweakdissipation}, though this does not change the qualitative details of the mechanism. We have also neglected thermal effects from the dissipation which assumes the dark gauge boson does not thermalise fast enough. It would be interesting to consider such thermal effects in both the dark and visible sectors and how they could be used more directly in our model~\footnote{See for example Refs.~\cite{hardy,hooktavares}.}.

\section{Conclusion}
\label{sec:conclusion}

We introduced an alternative cosmological relaxation model in which the backreaction of the Higgs vacuum expectation value determines the end of inflation rather than the end of relaxation. The relaxion is instead trapped indirectly as a result of Hubble falling below a certain critical threshold. The advantage of such a mechanism is to simplify the relaxion sector, which remains rather minimal, while overcoming difficulties in requiring the trapping to depend directly on the Higgs' backreaction. 

In this work we proposed two possible models of inflation in which a vacuum expectation value of the Higgs at the weak scale could trigger the end of inflation. In the first, the inflaton is slow-rolling due to dissipation into massless electroweak gauge bosons. As the vacuum expectation value switches on and the gauge bosons gain mass, this frictional force switches off and hence inflation ends. In the second model the inflaton is a QCD axion trapped by the barriers of its periodic potential on a larger hilltop potential slope, the latter originating from a larger periodic potential of a different gauge group. In this case the vacuum expectation value determines the height of the barrier, which starts out high enough to block the inflaton from rolling. At a small enough value near the weak scale the inflaton may then roll down the hilltop potential. 

We also briefly mentioned the minimal possibility of weak-scale hybrid inflation, with the relaxion as the inflaton and the Higgs as the waterfall field terminating inflation upon electroweak symmetry breaking. This scenario is more restrictive but remarkably economical in that it is already fulfilled by the relaxion setup and does not require a separate inflationary sector. 

Once inflation ends the universe becomes radiation-dominated and the Hubble scale decreases linearly in time. At some point the relaxion's axial gauge field coupling of a dark gauge group causes enough friction relative to Hubble that the relaxion loses the kinetic energy that enabled it to overcome the bumps in its periodic potential, and subsequently becomes trapped.

While this model shares many ingredients with previous realisations of cosmological relaxation, the origin of a naturally large hierarchy is subtly different. Previously the weak scale was due to a Higgs-dependent trapping backreaction, and the scanning need not even occur during inflation. On the contrary, here the weak scale is directly determined by the duration of a Higgs-dependent inflation -- specifically the exponential ratio of the scanning before and after inflation ends.

There are many avenues to explore in this direction. The general mechanism we have outlined here may be further developed into a more realistic relaxion and inflaton model. There could also be other possibilities to explore for inflationary backreactions. Moreover, quite independently of the hierarchy problem, it is interesting in its own right to consider various ways for triggering the end of inflation.

As experimental data may point towards no new physics at the weak scale, this motivates considering alternative ways in which backreaction in cosmological relaxation can dynamically select an apparently fine-tuned vacuum. More generally, we must fully explore scenarios of cosmological evolution that depend not only on the Higgs field but also on its vacuum expectation value. There are many studies of how the Higgs can play a key role during the early universe, but it may well be the early universe that is crucial to understanding the lightness of the Higgs.

\subsection*{Acknowledgements}
I thank Matthew McCullough for his collaboration, suggestions and encouragement. I am also grateful to Anson Hook and Ed Hardy for useful comments, Eugene Lim for related discussions, and the hospitality of CERN where part of this work was completed. I am supported by a Junior Research Fellowship from Gonville and Caius College, Cambridge.

\begin{appendix}

\end{appendix}

\newpage

 \providecommand{\href}[2]{#2}\begingroup\raggedright

\end{document}